\documentclass[twocolumn,preprintnumbers,endnote,nofootinbib,prl]{revtex4}
\usepackage{graphicx}
\usepackage{amsmath}

\usepackage[hypertex]{hyperref}
\usepackage[dvips]{color}

\newcommand{\lsim}{\lesssim}
\newcommand{\gsim}{\gtrsim}

\newcommand{\beq}{\begin{equation}}
\newcommand{\eeq}{\end{equation}}

\newcommand{\mP}{\bar{M}_{\rm P}}
\newcommand{\Lphi}{\Lambda_\phi}
\newcommand{\ktil}{\tilde{k}}
\newcommand{\invfb}{${\rm fb}^{-1}$}

\begin{document}

\pagestyle{plain}

\title{Precocious Diphoton Signals of the Little Radion at Hadron Colliders}

\author{Hooman Davoudiasl
}

\author{Thomas McElmurry
}

\author{Amarjit Soni
}
\affiliation{Department of Physics, Brookhaven National
Laboratory, Upton, NY 11973, USA}


\begin{abstract}

In Little Randall-Sundrum models, 
the bulk couplings of the radion to massless 
gauge fields can yield a greatly enhanced
diphoton signal at hadron colliders. We examine the implications of
the Tevatron data for the Little radion and also show that the 7~TeV run
at the Large Hadron Collider will have an impressive reach in this channel.  The
diphoton signal is crucial in the search for a light radion, or the
dual {\it dilaton}, and can potentially probe the ultraviolet scale of the
theory.

\end{abstract}
\maketitle


\underline{Introduction:} The 5D Randall-Sundrum (RS) model 
was initially proposed to explain the 
hierarchy between the 4D Planck mass $\mP\sim 10^{18}$~GeV 
and the weak scale $\sim 100$~GeV \cite{Randall:1999ee}.      
The necessary stabilization of the RS compact
dimension \cite{GW1} gives rise to a scalar, 
the radion $\phi$, with couplings governed by the warped-down
gravity scale \cite{GW2}.  Typically, $\phi$ is the lightest RS 
state \cite{GW2} and could provide one of the first glimpses of 
physics beyond the Standard Model (SM).  Placing the SM 
gauge fields \cite{Davoudiasl:1999tf,Pomarol:1999ad} and fermions
\cite{Grossman:1999ra} in the 5D RS bulk can result in an 
appealing framework for explaining the hierarchy and flavor puzzles
simultaneously \cite{Grossman:1999ra,Gherghetta:2000qt}.  In these models, 
myriad new resonances, the Kaluza-Klein (KK) excitations of the 
5D fields, appear at the TeV scale \cite{Davoudiasl:2009cd}.  However, discovery of 
such states would not  immediately establish that the underlying RS model 
has an ultraviolet (UV) cutoff scale near $\mP$; such a strong 
assumption would need experimental verification.  In fact,  
viewed primarily as a natural    
framework for describing SM flavor, 
the RS background may be governed by a UV scale 
far below $\mP$ \cite{LRS}.  In such 
volume-truncated ``Little Randall-Sundrum (LRS)" models,
some constraints can be alleviated and various KK signals can
be markedly enhanced \cite{LRS,LRSLHC,Rehermann:2010vq}.

The tree-level couplings of the original RS radion were proportional
to particle masses \cite{GW2,Giudice:2000av,radphen}. However, 
bulk fields have additional radion couplings; in particular,
the massless gluon $g$ and the photon $\gamma$ can couple to $\phi$
at the tree level \cite{Rizzo:2002pq,CHL}, with a strength inversely
proportional to the volume of the extra dimension. Hence, the
diphoton signal for $\phi$ can be significantly enhanced in
volume-truncated LRS models \cite{Toharia:2008tm}.  The clean diphoton signal provides the most
important search mode for the radion, below the $WW$ threshold, 
and even above the $WW$ threshold the enhanced
diphoton branching fraction can provide a diagnostic
handle to distinguish LRS models from other new physics
scenarios. For example, note that, while the gluon-fusion
production of the Higgs boson with four SM generations
can be enhanced by a factor of $\sim10$, the corresponding
diphoton branching fraction is actually reduced from the 
three-generation value to well below $10^{-3}$ \cite{Kribs:2007nz}.   
The sensitivity of the diphoton signal to the 5D volume can potentially 
yield information about the UV 
regime of the theory.  Also, the $\phi\gamma\gamma$
coupling is largely insensitive to
the content of the model above the radion mass $m_\phi$ and much of our
analysis can be applied to models with extended gauge symmetries
\cite{custodial}. Given these considerations, we focus in this work on the diphoton
signal for the Little radion at hadron colliders.
We will examine Little radion phenomenology at the Tevatron and also
show that the early 7~TeV run at the 
Large Hadron Collider (LHC) has a significant reach
for interesting values of parameters.  

\underline{Formalism and setup:} The RS background is a slice
of 5D anti-de~Sitter (AdS) spacetime with the metric
$ds^2 = e^{-k y} \eta_{\mu \nu}d x^\mu d x^\nu - dy^2$ \cite{Randall:1999ee},
where $k$ is the curvature scale, typically
assumed smaller than the
5D fundamental  scale $M_5$.  The compact dimension $y$ is
bounded by UV and infrared (IR) branes at $y=0,L$, respectively.
We will assume that electroweak
symmetry is broken by an IR-brane-localized
Higgs doublet.  In principle, there can be
other operators, such as kinetic terms, localized on the
branes.  However, for simplicity, we ignore such terms;
this assumption can be justified since these operators naturally
arise as counterterms for regulating bulk quantum loops \cite{Georgi:2000ks}
and can be expected to have small coefficients.
The RS framework for SM flavor is based on the localization of
bulk fermion zero modes along $y$, towards or away from
the IR-localized Higgs.  The localization of a bulk fermion $\Psi$ depends
on its 5D mass $m_\Psi$, often parameterized by $c_\Psi \equiv m_\Psi/k$.

The radion is associated with the quantum fluctuations
in the size $L$ of the fifth dimension.  The 
couplings of $\phi$ to the trace of the energy-momentum tensor
are suppressed by the scale $\Lphi \equiv \lambda \sqrt{6 M_5^3/k}$
\cite{GW2}, where $\lambda\equiv e^{-k L}$. We now briefly 
review the radion interactions with bulk fields,
detailed in Ref.~\cite{CHL}. The coupling of $\phi$ to SM vector field
$V^{(i)}$, $i=W,Z$, of mass $m_i$ is given by
\beq {\cal L}_V =
-\frac{\phi}{\Lphi}\sum_{i} a_i \left[\mu_i^2 \, V^{(i)}_\mu
V^{(i)\,\mu} + \frac{1}{4 \, kL}
V^{(i)}_{\mu\nu}V^{(i)\,\mu\nu}\right],
\label{phiVV}
\eeq where
$V_{\mu\nu} \equiv \partial_\mu V_\nu - \partial_\nu V_\mu$, $a_{W}
= 2,\,a_{Z}=1$, and 
\beq \mu_i^2 = m_i^2 \left[1 - \frac{kL}{2} \left(\frac{m_i}{\tilde{k}}\right)^2\right] \,.
\label{mui2} 
\eeq 
Here,  $\tilde{k}\equiv k \lambda$ sets the scale
of the lightest KK masses and corrections to ${\cal L}_V$ are
suppressed by powers of $m_i^2/{\tilde{k}}^2$.

A massless gauge field $A_\mu$ couples to $\phi$ via
\beq {\cal L}_A= -\frac{\phi}{4 \Lphi\, kL} \left[1 + \frac{\alpha_G}{2
\pi} \, b_G\, kL\right]F_{\mu\nu}F^{\mu\nu},
\label{phiAA}
\eeq
where $b_G$ denotes the 1-loop $\beta$-function coefficient
below $m_\phi/2$ and $F_{\mu\nu}\equiv \partial_\mu A_\nu -
\partial_\nu A_\mu$.  In this work, the relevant gauge
fields are $\gamma$ ($G=\text{EM}$) and $g$
($G=\text{QCD}$). We have
\beq b_{\rm EM}= b_2 + b_Y - F_W -
\frac{4}{3} F_t,
\label{bem}
\eeq
with $b_2 = 19/6$, $b_Y = -41/6$.  To a good approximation, $F_W = 7$ for $m_\phi <  2 m_W$, 
$F_t = -4/3$ for $m_\phi < 2 m_t$, and both functions are zero
when $\phi$ is heavier than twice the mass of the respective
particle.  For gluons, \beq b_{\rm QCD}= 11 - \frac{2}{3} N_F,
\label{bqed} \eeq where $N_F$ is the number of quark flavors.

The coupling of $\phi$ to SM fermion $f$ of mass $m_f$ depends on its bulk
profile parameters $c_{L,R}$, corresponding to the left and right
4D chiralities, respectively \cite{CHL,ATZ}:
\beq -\frac{\phi}{\Lphi} m_f [I(c_L) + I
(c_R)]\, ({\bar f_L} f_R + {\bar f_R} f_L),
\label{phiff}
\eeq
where
\beq
I(c) \equiv \frac{1 - 2 c}{2\,(1 - \lambda^{1-2c})} + c\,.
\label{I(c)}
\eeq
In our convention, fermions with $c_{L,R}>1/2$ have UV-localized zero modes.

Finally, the coupling of the radion to a brane-localized Higgs
scalar is given by~\cite{Giudice:2000av} \beq
-\frac{\phi}{\Lphi}(-\partial_\mu h \partial^\mu h + 2\, m_h^2\,
h^2), \label{phihh} \eeq where $h$ is the physical Higgs of mass
$m_h$. We will next discuss the relevant LRS parameters for our
analysis.

Let us begin with $m_\phi$, which
depends on the radius stabilization potential.  For concreteness and
as a guide for our phenomenological study, we will
consider the Goldberger-Wise (GW) mechanism
\cite{GW1,GW2}, where a bulk scalar $\Phi$ of mass $m$ with brane-localized
potentials is introduced.  Let us denote the 5D vacuum expectation
value of $\Phi$ on the UV and the IR branes by  $v_0$ and
$v_L$ (with mass dimension 3/2), respectively. Then, one can show
that the stabilized radius $L$ is given by \cite{GW1,GW2}
\beq
k L = \epsilon^{-1} \ln(v_0/v_L),
\label{GWkL}
\eeq
where
$\epsilon\equiv m^2/(4 k^2)$ and
\beq
m_\phi^2 =
\frac{v_L^2}{3 M_5^3}\, \epsilon^2 \tilde{k}^2.
\label{GWmphi}
\eeq

To simplify our treatment, we henceforth set $k=M_5$ which implies $\Lphi
= \sqrt{6}\, \ktil$.  Also, since $v_0$
and $v_L$ are 5D parameters, it is reasonable to assume that
$v_{0,L}\sim k^{3/2}$ and $ \ln(v_0/v_L) \sim 1$, which yields $\epsilon
\sim (k L)^{-1}$, from Eq.(\ref{GWkL}). Using Eq.(\ref{GWmphi}), we
then get $m_\phi \sim \tilde{k}/(k L)$. Various constraints suggest
$\tilde{k}\gsim 1$~TeV \cite{custodial}.  Here, we will typically
assume $kL \sim 7$, as a lower bound for realistic models
of flavor allowed by precision data
\cite{LRSflavor}.  Hence, $m_\phi \sim 100{\rm~GeV}$ is a
reasonable choice, which we will adopt for the Little radion in this
work. For definiteness, we will set $kL=7$ and
use the LRS fermion profiles presented in
Ref.~\cite{LRSflavor}.

\underline{Results:}
\begin{figure}
\includegraphics[width=0.48\textwidth]
{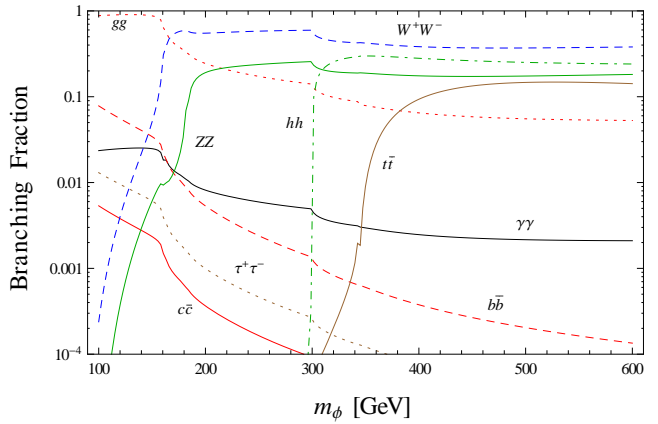}
\caption{Little radion branching fractions, for $kL=7$ and $\Lphi = 3$~TeV.}
\label{fig:branch}
\end{figure}
Various branching fractions of the Little radion, over
a wide range of $m_\phi$, for $\Lphi=3$~TeV, are presented in Fig.~\ref{fig:branch}.  Contributions from
fermions lighter than the charm quark are ignored in our analysis.  For definiteness, 
we assume $m_h=150$~GeV in this work, unless otherwise specified.

The qualitative features in Fig.~\ref{fig:branch} 
are similar to those in the conventional
bulk-field scenarios \cite{CHL,MT}.   However, we note that 
the $\gamma\gamma$ branching fraction is $\sim 1\%$,
enhanced by about an order of magnitude compared to the
conventional prediction.  This is due to the larger tree-level
$\phi \gamma\gamma$ coupling (inversely proportional to $kL$) in our 
LRS setup.  

Note also that the width 
$\Gamma_\phi$ of $\phi$, which approximately  
scales as $\Lphi^{-2}$, is quite narrow: 
$\Gamma_\phi \approx 0.4$~MeV (1.1~GeV) 
for $\Lphi=3$~TeV and $m_\phi=100$ (600)~GeV.  
For comparison, the SM Higgs width is  
$\Gamma_h \approx 2.5$~MeV (123~GeV), 
for $m_h=100$ (600)~GeV \cite{Hwidth}. 

At hadron colliders, our signal process is
$gg\to\phi\to\gamma\gamma$, whose main background is SM diphoton
production. To compute this background, we use DIPHOX
\cite{Binoth:1999qq}, which includes the partonic subprocess $q\bar
q\to\gamma\gamma$ through next-to-leading order, as well as 
$gg\to\gamma\gamma$, which is formally
next-to-next-to-leading order but
numerically significant. We neglect the interference 
of the latter with the signal process and note that its analog in the case of
Higgs production has been shown to be small \cite{Dicus:1987fk,
Dixon:2003yb}.  We use
the CUBA library \cite{Hahn:2004fe} in the computation of the signal,
and the CTEQ6M parton distribution functions \cite{Pumplin:2002vw} for
both signal and background.

\begin{figure}
\includegraphics[width=0.48\textwidth]
{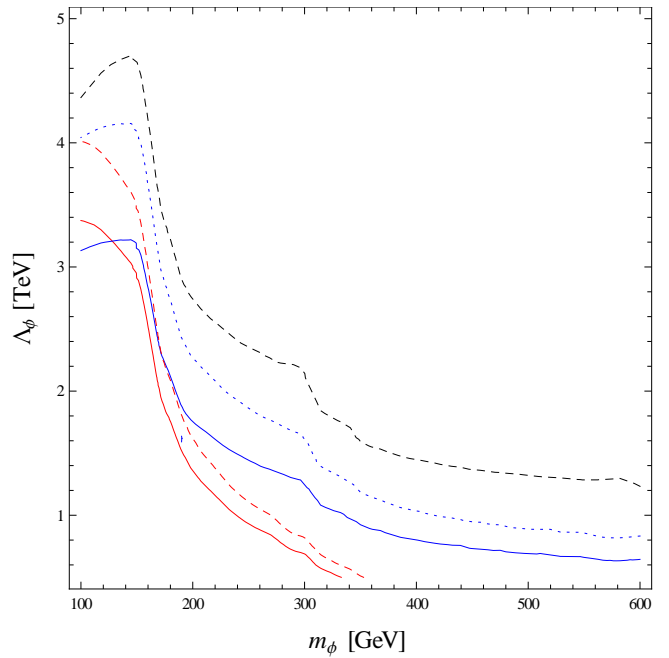} \caption{Reach in the $\gamma\gamma$ channel for
$(m_\phi, \Lphi)$, with $kL=7$.  The lower curves are the $2\sigma$
contours for the Tevatron with 5~\invfb\ (solid) and 10~\invfb\
(dashed).  The middle curves represent the $5\sigma$ (solid) and
$3\sigma$ (dotted) contours for the LHC with $\sqrt{s}=7$~TeV and
1~\invfb.  The uppermost contour marks the $5\sigma$ reach at the
LHC with $\sqrt{s}=14$~TeV and 1~\invfb.} \label{fig:reach}
\end{figure}

The collider reach for the LRS radion $\phi$, in the diphoton
channel, is presented in Fig.~\ref{fig:reach}.  Here, the lower
curves correspond to the $2\sigma$ reach of Run II at the Tevatron
with 5~\invfb\ (solid) and 10~\invfb\ (dashed). We have assumed a
70\% efficiency for photon identification \cite{D0photon,CDFRSgrav}.
We impose cuts of $p_T>25$~GeV and $|\eta|<1.1$ on the
transverse momentum $p_T$ and rapidity $\eta$ of each photon,
following Refs.~\cite{D0RSgrav,CDFRSgrav}. We require the two
photons to be separated from one another by $\Delta
R\equiv\sqrt{(\Delta\varphi)^2+(\Delta\eta)^2}>0.4$ ($\varphi$ is
the azimuthal angle), and that the total hadronic transverse energy
within a cone of radius $\Delta R=0.4$ around each photon be less
than 10 GeV. Finally, we impose an invariant mass cut, requiring
$|m_{\gamma\gamma}-m_\phi|<10$~GeV. At the $2\sigma$ level
of significance, the Tevatron can probe $\Lphi \approx 1$~TeV and
$m_\phi \lsim 300$~GeV, as well as $m_\phi\approx 100$~GeV and
$\Lphi \lsim 2.5$~TeV, in the $\gamma \gamma$ channel. Currently,
the CDF collaboration reports an excess of $\sim 10$ events above
the SM background for $m_{\gamma\gamma}\approx 200$~GeV with
5.4~\invfb\ \cite{CDFRSgrav}.  We find that the Little radion with
$kL\approx 7$ can potentially produce this level of excess for
$\Lphi \approx 1$~TeV at $m_\phi \approx 200$~GeV. Again with
5.4~\invfb, the D0 collaboration finds a $2.3\sigma$ deviation from
background at $m_{\gamma\gamma}\approx 450$~GeV \cite{D0RSgrav}. A
Little radion signal at this level would require $kL \approx 7$ and
$\Lphi < 500$~GeV, which is not easily accommodated in typical
warped flavor models.  Apart from these two anomalies, one may
consider the parameter space below the 5~\invfb\ contour to be
disfavored, based on the current Tevatron data.

We now examine the diphoton signal of the Little radion at
the LHC, assuming a photon identification efficiency of 80\%
\cite{LHCphoton}. We apply the cuts $p_T>30$~GeV and
$|\eta|<2.4$, similar to those used for Higgs searches in
Ref.~\cite{:1999fr}, and the same isolation and invariant mass cuts
as above. The middle curves in Fig.~\ref{fig:reach} represent the
$5\sigma$ (solid) and $3\sigma$ (dotted) contours for the LHC with
$\sqrt{s}=7$~TeV and 1~\invfb, corresponding to the planned early
run.  We see that the $5\sigma$ reach is bracketed
by [$m_\phi\approx 100$~GeV,  $\Lphi\approx 3$~TeV] and 
[$m_\phi\approx 300$~GeV,  $\Lphi\approx 1$~TeV].  
Thus, the diphoton signal of the Little radion can be conclusively detected
during the early LHC run, over a wide range of parameters for
realistic LRS models of flavor.  The uppermost curve in
Fig.~\ref{fig:reach} marks the $5\sigma$ reach at the LHC
with $\sqrt{s}=14$~TeV and 1~\invfb, extending to 
$\Lphi \gsim 2$~TeV 
for $m_\phi \lsim 300$~GeV.

\underline{Duality:} The AdS/Conformal Field Theory correspondence 
\cite{ADSCFT} suggests that a 5D warped model is  
dual to a 4D theory with a conformal symmetry, dynamically emerging below a  
scale $\sim k$ and spontaneously broken at 
a scale $\sim \ktil$.  The geometric boundaries are then identified with the
the UV and IR scales of the 4D dynamics \cite{holography}. 
This holographic picture implies that $\phi$ is dual to a dilaton arising from the
spontaneous breaking of conformal symmetry at the IR scale, and can
be expected to be light; some recent work on various aspects 
of light dilaton physics can be found in 
Refs.~\cite{Goldberger:2007zk,Fan:2008jk,Appelquist:2010gy}.

The phenomenology of the radion and that of the dilaton are  
qualitatively very similar.  However, 5D
warped models represent a specific realization of 4D dynamics, giving 
rise to features that are absent in general discussions.
In particular, tree-level couplings of $\phi$ to massless gauge fields, proportional
to $1/(kL)$ [see Eq.~(\ref{phiAA})], do not have 
counterparts among the dilaton couplings from 
Refs.~\cite{Goldberger:2007zk,Fan:2008jk}.  This can
be understood  by noting that the absence of a UV cutoff for
the 4D conformal dynamics implies $kL \to \infty$,
which removes the tree-level couplings \cite{YB}.  
Hence, a truncated LRS volume corresponds to
a reduced {\it conformal depth} beyond the weak scale for the 4D dynamics.
Then, to see the enhanced coupling of the Little dilaton to massless gauge fields,
one presumably needs to include the effects of a
4D UV cutoff, below which the theory flows to conformal dynamics.

\underline{Diagnostics:} 
Within our framework, a diphoton signal similar to those discussed 
here would favor the LRS version of warped models.  In addition, 
Eq.~(\ref{phiVV}) implies 
that, for $m_\phi \lsim 400$~GeV, the branching 
fractions into massive vectors are dominated by 
the IR-brane-localized couplings of $\phi$ proportional to $m_{W, Z}^2$.  
Thus, for a wide range of $m_\phi \gsim 2 m_W$, one can extract an estimate 
of $kL$, by comparing the cross sections in the 
$\gamma\gamma$, $W W$, and $ZZ$ channels.  This information 
will provide more direct evidence that the relevant warped scenario is of the LRS type, with a 
UV scale far below $\mP$.  In the 4D dual description, this hints at the UV scale where 
conformal behavior emerges.  Further confirmation 
of this picture can be obtained by detecting clean LRS signals associated with 
the TeV-scale resonances (KK modes) \cite{LRS,LRSLHC}, perhaps at later experimental stages.

\underline{Conclusions:}
In this work, we examined the clean diphoton signal of the radion 
in Little Randall-Sundrum models, at hadron colliders.  
These volume-truncated models explain SM flavor in a geometric fashion, and maintain
a hierarchy between the weak scale and a large sub-Planckian scale.  
We considered a UV scale of order a few thousand TeV, 
corresponding to $kL=7$, for viable models of flavor.
The enhanced couplings of the Little radion in this framework afford
a much larger diphoton signal at the Tevatron and the LHC.
We commented on the implied bounds and 
potential LRS candidates for some of
the observed anomalies, based on current Tevatron data.  
The initial 7 TeV run at the LHC can decisively 
detect the diphoton signal of the Little radion, over 
a wide range of parameters, with
1~\invfb\ of integrated luminosity.  The sensitive dependence of this radion signal on the LRS
volume can potentially yield information on the underlying  
UV cutoff scale of the model, early on.  In
a dual 4D description, this can be interpreted as information on 
the UV scale where the dynamics becomes conformal.


\acknowledgments

We thank Y.~Bai, S.~Dawson, J.~Hubisz, and M.~Toharia for discussions, and J.-P.~Guillet for assistance with DIPHOX.
This work is supported in part by the US DOE
Grant DE-AC02-98CH10886.


{\it Note added:} In an earlier version of this work, the branching 
fractions for $WW$ and $ZZ$, plotted as a function of $m_\phi$  in Fig.~\ref{fig:branch},  
sharply decreased near certain $m_\phi$ values, due to an error in our computation of these curves.  
We have corrected this error in the current version; our results now agree with those presented in 
Ref.~\cite{Barger:2011qn}.  We have also implemented the appropriate 
changes in Fig.~\ref{fig:reach} and parts of the text.  An erratum reflecting the aforementioned corrections 
will be prepared for the published version of this paper \cite{Davoudiasl:2010fb}.  
Our main conclusions about the enhanced diphoton 
signals of the Little radion remain qualitatively the same and quantitatively similar to the previous results.  



\begin{thebibliography}{99}

\bibitem{Randall:1999ee}
  L.~Randall and R.~Sundrum,
  Phys.\ Rev.\ Lett.\  {\bf 83}, 3370 (1999)
  [arXiv:hep-ph/9905221].

\bibitem{GW1}
  W.~D.~Goldberger and M.~B.~Wise,
  Phys.\ Rev.\ Lett.\  {\bf 83}, 4922 (1999)
  [arXiv:hep-ph/9907447].

\bibitem{GW2}
  W.~D.~Goldberger and M.~B.~Wise,
  Phys.\ Lett.\  B {\bf 475}, 275 (2000)
  [arXiv:hep-ph/9911457].

\bibitem{Davoudiasl:1999tf}
  H.~Davoudiasl, J.~L.~Hewett and T.~G.~Rizzo,
  Phys.\ Lett.\  B {\bf 473}, 43 (2000)
  [arXiv:hep-ph/9911262];

\bibitem{Pomarol:1999ad}
  A.~Pomarol,
  Phys.\ Lett.\  B {\bf 486}, 153 (2000)
  [arXiv:hep-ph/9911294].

\bibitem{Grossman:1999ra}
  Y.~Grossman and M.~Neubert,
  Phys.\ Lett.\  B {\bf 474}, 361 (2000)
  [arXiv:hep-ph/9912408].

\bibitem{Gherghetta:2000qt}
  T.~Gherghetta and A.~Pomarol,
  Nucl.\ Phys.\  B {\bf 586} (2000) 141
  [arXiv:hep-ph/0003129].

\bibitem{Davoudiasl:2009cd}
  For a recent review, see, for example: 
H.~Davoudiasl, S.~Gopalakrishna, E.~Pont\'on and J.~Santiago,
  New J.\ Phys.\  {\bf 12}, 075011 (2010)
  [arXiv:0908.1968 [hep-ph]].

\bibitem{LRS}
  H.~Davoudiasl, G.~Perez and A.~Soni,
  Phys.\ Lett.\  B {\bf 665}, 67 (2008)
  [arXiv:0802.0203 [hep-ph]].

\bibitem{LRSLHC}
  H.~Davoudiasl, S.~Gopalakrishna and A.~Soni,
  Phys.\ Lett.\  B {\bf 686}, 239 (2010)
  [arXiv:0908.1131 [hep-ph]].

\bibitem{Rehermann:2010vq}
  K.~Rehermann and B.~Tweedie,
  arXiv:1007.2221 [hep-ph].

\bibitem{Giudice:2000av}
G.~F.~Giudice, R.~Rattazzi and J.~D.~Wells,
  Nucl.\ Phys.\  B {\bf 595}, 250 (2001)
  [arXiv:hep-ph/0002178];

\bibitem{radphen}
Early work on radion physics can also be found in:  C.~Cs\'aki, M.~L.~Graesser and G.~D.~Kribs,
  Phys.\ Rev.\  D {\bf 63}, 065002 (2001)
  [arXiv:hep-th/0008151];
  J.~L.~Hewett and T.~G.~Rizzo,
  JHEP {\bf 0308}, 028 (2003)
  [arXiv:hep-ph/0202155];
  J.~F.~Gunion, M.~Toharia and J.~D.~Wells,
  Phys.\ Lett.\  B {\bf 585}, 295 (2004)
  [arXiv:hep-ph/0311219].

\bibitem{Rizzo:2002pq}
  T.~G.~Rizzo,
  JHEP {\bf 0206}, 056 (2002)
  [arXiv:hep-ph/0205242].

\bibitem{CHL}
  C.~Cs\'aki, J.~Hubisz and S.~J.~Lee,
  Phys.\ Rev.\  D {\bf 76}, 125015 (2007)
  [arXiv:0705.3844 [hep-ph]].

\bibitem{Toharia:2008tm} 
For a study of enhanced diphoton signals with radion-
Higgs mixing in the original RS model, see M.~Toharia,
  Phys.\ Rev.\ D {\bf 79}, 015009 (2009)
  [arXiv:0809.5245 [hep-ph]].


\bibitem{Kribs:2007nz} 
  G.~D.~Kribs, T.~Plehn, M.~Spannowsky and T.~M.~P.~Tait,
  Phys.\ Rev.\ D {\bf 76}, 075016 (2007)
  [arXiv:0706.3718 [hep-ph]].


\bibitem{custodial}
  K.~Agashe, A.~Delgado, M.~J.~May and R.~Sundrum,
  JHEP {\bf 0308}, 050 (2003)
  [arXiv:hep-ph/0308036];
  K.~Agashe, R.~Contino, L.~Da Rold and A.~Pomarol,
  Phys.\ Lett.\  B {\bf 641}, 62 (2006)
  [arXiv:hep-ph/0605341];
  M.~S.~Carena, E.~Pont\'on, J.~Santiago and C.~E.~M.~Wagner,
  Phys.\ Rev.\  D {\bf 76}, 035006 (2007)
  [arXiv:hep-ph/0701055].

\bibitem{Georgi:2000ks}
  H.~Georgi, A.~K.~Grant and G.~Hailu,
  Phys.\ Lett.\  B {\bf 506}, 207 (2001)
  [arXiv:hep-ph/0012379].

\bibitem{ATZ}
  A.~Azatov, M.~Toharia and L.~Zhu,
  Phys.\ Rev.\  D {\bf 80}, 031701 (2009)
  [arXiv:0812.2489 [hep-ph]].

\bibitem{LRSflavor}
  M.~Bauer, S.~Casagrande, L.~Gr\"under, U.~Haisch and M.~Neubert,
  Phys.\ Rev.\  D {\bf 79}, 076001 (2009)
  [arXiv:0811.3678 [hep-ph]].


\bibitem{MT}
Contribution by M.~Toharia in:
  P.~Nath {\it et al.},
  Nucl.\ Phys.\ Proc.\ Suppl.\  {\bf 200-202}, 185 (2010)
  [arXiv:1001.2693 [hep-ph]].

\bibitem{Hwidth}
{\tt https://twiki.cern.ch/twiki/bin/view/LHCPhysics/}\\
{\tt CrossSections.}

\bibitem{Binoth:1999qq}
  T.~Binoth, J.~P.~Guillet, E.~Pilon and M.~Werlen,
  Eur.\ Phys.\ J.\  C {\bf 16}, 311 (2000)
  [arXiv:hep-ph/9911340].

\bibitem{Dicus:1987fk}
  D.~A.~Dicus and S.~S.~D.~Willenbrock,
  Phys.\ Rev.\  D {\bf 37}, 1801 (1988).

\bibitem{Dixon:2003yb}
  L.~J.~Dixon and M.~S.~Siu,
  Phys.\ Rev.\ Lett.\  {\bf 90}, 252001 (2003)
  [arXiv:hep-ph/0302233].

\bibitem{Hahn:2004fe} 
  T.~Hahn,
  Comput.\ Phys.\ Commun.\  {\bf 168}, 78 (2005)
  [hep-ph/0404043].


\bibitem{Pumplin:2002vw}
  J.~Pumplin, D.~R.~Stump, J.~Huston, H.~L.~Lai, P.~M.~Nadolsky and W.~K.~Tung,
  JHEP {\bf 0207}, 012 (2002)
  [arXiv:hep-ph/0201195].

\bibitem{CDFRSgrav}
Search for Randall-Sundrum Gravitons in the Diphoton Channel at CDF, Public Note,
The CDF Collaboration, http://www-cdf.fnal.gov.

\bibitem{D0photon}
The D0 Collaboration, D0 note 5369-CONF.

\bibitem{D0RSgrav}
  V.~M.~Abazov {\it et al.}  [The D0 Collaboration],
  Phys.\ Rev.\ Lett.\  {\bf 104}, 241802 (2010)
  [arXiv:1004.1826 [hep-ex]].

\bibitem{LHCphoton}
Contribution by L.~R.~Flores Castillo in:
  P.~Nath {\it et al.},
  Nucl.\ Phys.\ Proc.\ Suppl.\  {\bf 200-202}, 185 (2010)
  [arXiv:1001.2693 [hep-ph]].

\bibitem{:1999fr}
 ``ATLAS detector and physics performance. Technical design report.  Vol. 2.''





\bibitem{ADSCFT}
  J.~M.~Maldacena,
  Adv.\ Theor.\ Math.\ Phys.\  {\bf 2}, 231 (1998)
  [Int.\ J.\ Theor.\ Phys.\  {\bf 38}, 1113 (1999)]
  [arXiv:hep-th/9711200].

\bibitem{holography}
  See, for example, N.~Arkani-Hamed, M.~Porrati and L.~Randall,
  JHEP {\bf 0108}, 017 (2001)
  [arXiv:hep-th/0012148];
  R.~Rattazzi and A.~Zaffaroni,
  JHEP {\bf 0104}, 021 (2001)
  [arXiv:hep-th/0012248].

\bibitem{Goldberger:2007zk}
  W.~D.~Goldberger, B.~Grinstein and W.~Skiba,
  Phys.\ Rev.\ Lett.\  {\bf 100}, 111802 (2008)
  [arXiv:0708.1463 [hep-ph]].

\bibitem{Fan:2008jk}
  J.~Fan, W.~D.~Goldberger, A.~Ross and W.~Skiba,
  Phys.\ Rev.\  D {\bf 79}, 035017 (2009)
  [arXiv:0803.2040 [hep-ph]].

\bibitem{Appelquist:2010gy}
  T.~Appelquist and Y.~Bai,
  arXiv:1006.4375 [hep-ph].

\bibitem{YB} 
We thank Y. Bai for discussions on this point.

\bibitem{Barger:2011qn} 
  V.~Barger and M.~Ishida,
  Phys.\ Lett.\ B {\bf 709}, 185 (2012)
  [arXiv:1110.6452 [hep-ph]].

\bibitem{Davoudiasl:2010fb} 
  H.~Davoudiasl, T.~McElmurry and A.~Soni,
  Phys.\ Rev.\ D {\bf 82}, 115028 (2010)
  [arXiv:1009.0764 [hep-ph]].



\end{thebibliography}
\end{document}